\documentclass[journal]{IEEEtran}
\usepackage{amsmath,amssymb}
\usepackage{graphicx}
\usepackage{booktabs}
\usepackage{array}
\usepackage{siunitx}
\usepackage{xcolor}
\usepackage[hidelinks]{hyperref}
\usepackage{stfloats}          

\hypersetup{pdftitle={Wideband AR-HCF Transmission Under Variable Network Traffic: A System-Level Analysis},
  pdfauthor={Md Ghulam Saber and Zhiping Jiang},
  pdfsubject={Hollow-core fiber, energy-efficient elastic optical networks},
  pdfkeywords={hollow-core fiber, gas absorption, energy efficiency, ISRS, GN model}}
\graphicspath{{../figures_v3/}{./figures_v3/}}
\DeclareSIUnit{\dbm}{dBm}
\DeclareSIUnit{\baud}{Bd}
\DeclareSIUnit{\tbps}{Tb/s}   
\newcommand{\gsnr}{\mathrm{GSNR}}
\newcommand{\dbkm}{\decibel\per\kilo\meter}

\begin{document}

\title{Wideband AR-HCF Transmission Under Variable Network Traffic: A System-Level Analysis}
\author{Md Ghulam Saber and Zhiping Jiang%
	\thanks{M. G. Saber and Z. Jiang are with the Ottawa Research Center, Huawei Technologies Canada, 303 Terry Fox Drive, Kanata, ON, K2K 3J1,  Canada. e-mail:md.ghulam.saber@huawei.com.}%
}

	\markboth{Journal of \LaTeX\ Class Files,~Vol.~xx, No.~x, xxx~2026}%
{Shell \MakeLowercase{\textit{et al.}}: Bare Demo of IEEEtran.cls for IEEE Journals}

\maketitle

\begin{abstract}

	Dynamic traffic changes spectral occupancy in wideband systems, perturbing surviving channels in conventional single-mode fiber (SMF) through Kerr nonlinear interference (NLI), inter-channel stimulated Raman scattering (ISRS), and amplifier loading, which motivates holding full occupancy with amplified-spontaneous-emission (ASE) idlers. Whether that requirement persists in antiresonant hollow-core fiber (AR-HCF) is examined with a span-by-span model of a 96-channel, 9.6-THz extended-C+L link over 960~km that jointly propagates signal, ASE, and NLI while accounting for amplifier pigtails, line-resolved CO$_2$ absorption, intermodal interference, amplifier control, and load transients. At a common launch power, changing occupancy from 100\% to 20\% shifts a surviving channel's generalized signal-to-noise ratio (GSNR) by only $1.6\times10^{-3}$~dB in AR-HCF against 1.52~dB in SMF, strongly reducing the steady-state motivation for spectral filling. Exploiting the AR-HCF headroom raises the throughput-optimal launch to +12.3~dBm/channel, where the same occupancy change shifts the surviving-channel GSNR by 0.24~dB; nearly all of this shift, and of the total NLI, arises in the short solid-glass amplifier pigtails---almost entirely the output pigtail. The pigtails hold the optimum launch 8.9~dB below the amplifier ceiling, yet removing them would add only about 1\% throughput at that launch, as the link is already near its transceiver, intermodal-interference, and gas-absorption limits. Line-resolved gas modeling separates a physical CO$_2$ attenuation of a few decibels, largely restored by per-span equalization, from the narrow in-band notches that flat amplifier gain cannot recover. As expected from amplifier physics, elastic operation saves 23\% of the amplifier energy under constant-gain control, which lets pump power follow the load, and none under constant-output control; an 85\% channel drop still produces 8.8~dB of uncontrolled gain excursion, reduced to 0.6~dB by an assumed 500-Hz control loop. The limits of elastic wideband transmission therefore shift from the fiber to amplifier-site components, gas absorption, and amplifier-control dynamics. 

\end{abstract}


\section{Introduction}

\IEEEPARstart{W}{ideband} optical networks must carry time-varying traffic while meeting stringent capacity and energy-efficiency requirements~\cite{uwb_energy2026}. At the physical layer the operating state of a conventional SMF link depends on its spectral occupancy: adding or dropping channels changes Kerr NLI, ISRS, and amplifier loading, and hence the generalized signal-to-noise ratio (GSNR) of the survivors~\cite{mitra2020fillmargin}. To stabilize performance, systems commonly hold an approximately constant load using ASE idlers~\cite{aseidler2024} or reconfigure the amplifiers as occupancy changes~\cite{song2022ampconfig}, but at an energy cost: the line-system power then stays near its full-load value however little traffic is carried, so consumption is largely decoupled from utilization.

AR-HCF offers a different propagation regime. Light is guided predominantly in air~\cite{fokoua2023loss}, giving a Kerr nonlinearity orders of magnitude below silica~\cite{hcf_kerr2023,hahn2026lpm}, losses approaching or below conventional fiber~\cite{hcf_sub01db2025}, and substantially greater launch-power headroom~\cite{hong2026tpo}. The reduced inter-channel coupling suggests an AR-HCF channel may be nearly insensitive to which neighbors are present, raising a system-level question: does a wideband AR-HCF line still require artificial spectral filling when traffic changes?

The answer is not determined by the transmission fiber alone. Practical AR-HCF links contain short solid-silica sections at amplifier sites carrying the full output power, which as launch power is raised to exploit the nonlinear headroom can become the dominant source of Kerr NLI. Hybrid-span experiments show the nonlinear threshold set by the solid-fiber portion of an otherwise hollow-core link~\cite{mardoyan2026hybridspan}, and longitudinal monitoring localizes nonlinear contributions at amplifier-connected solid sections~\cite{hahn2026lpm}, so the bottleneck may shift rather than disappear.

A second impairment, specific to AR-HCF, works against these gains: residual gas absorption. Narrow rotational-vibrational lines, particularly of carbon dioxide, fall inside coherent channels and add both excess loss and spectral distortion~\cite{ge2025fieldtrial,wang2025co2}. Because these features are much narrower than the signal bandwidth~\cite{hcf_gasline2026}, average attenuation cannot represent them: the excess loss raises the gain each amplifier must supply, and hence its ASE and pump power, while the narrow in-band notches distort the channel in a way amplification cannot undo. Gas absorption is thus a direct cost of the air-filled core that makes low-nonlinearity, idler-free operation possible, bearing on both the energy freed by removing idlers and the reach of the link. It motivates physical gas removal~\cite{xiong2025co2elim} and digital mitigation by spectral pre-equalization and sub-carrier multiplexing~\cite{hcf_gasline2026,wang2025co2}.

Recent studies have examined hollow-core energy efficiency and system limits: launch-power and link-energy optimization~\cite{sohanpal2026energyperbit,uwb_energy2026}, and limits set jointly by ASE, nonlinearity, IMI, and gas absorption~\cite{hcf_qamlimits2026}; spectral-loading penalties and dynamic amplifier reconfiguration have likewise been studied for elastic SMF systems~\cite{mitra2020fillmargin,song2022ampconfig}, and the consequences of constant-gain versus constant-output amplifier control under channel add/drop are well established~\cite{zyskind2011book,srivastava1997transient}. No prior work has assessed a wideband AR-HCF line as a single system under variable traffic, jointly accounting for propagation, amplifier-site nonlinearity, gas absorption, energy consumption, and transients. That gap is addressed here with a span-by-span model of a 96-channel extended-C$+$L link over \SI{960}{\kilo\meter} that propagates signal, ASE, and NLI together while accounting for occupancy-dependent ISRS, solid-glass amplifier pigtails, line-resolved CO$_2$ absorption, IMI, dynamic gain equalization, amplifier energy, and gain transients, with AR-HCF and G.652 SMF evaluated under identical loading to isolate the effect of the medium.

\section{System Model}
\label{sec:model}

\subsection{Channel Plan and Link}
The model uses \num{96} channels at \SI{96}{\giga\baud} on a
\SI{100}{\giga\hertz} grid: extended C and L each hold \num{48} channels over
\SI{4.8}{\tera\hertz}, amplified separately and separated by a
\SI{400}{\giga\hertz} guard, over $N_s=\num{12}$ spans of
\SI{80}{\kilo\meter}. Standard G.652
SMF and air-filled AR-HCF use the same line-system architecture; Table~\ref{tab:params}
lists the parameters.


\subsection{Nonlinear Interference}
\label{sec:nli}
Span NLI follows the closed-form ISRS-GN model~\cite{semrau2019isrsgn},
$P_{\mathrm{NLI},i}=[\eta_{\mathrm{SPM}}(f_i)+\eta_{\mathrm{XPM}}(f_i)]P_i^3$,
whose coefficients depend on the total power and the active neighbors, and hence
on occupancy. Self-channel NLI accumulates over spans with the coherence
exponent $\varepsilon$ of~\cite{poggiolini2012gn}, and each span's contribution
is referred to the loss-compensated plane. For AR-HCF, $\gamma$ follows from the
bound $n_2<\SI{2.2e-23}{\meter\squared\per\watt}$~\cite{hcf_kerr2023} and a
\SI{21}{\micro\metre} mode-field diameter, and the silica Raman slope is scaled
by an assumed $10^{-4}$ power fraction in glass~\cite{fokoua2023loss}, tested in
Sec.~\ref{sec:load}.

\subsection{Amplifier Pigtails}
Each amplifier site has a \SI{10}{\meter} SMF pigtail before and after the
amplifiers, both carrying the full C$+$L comb; with the equalizer co-located,
these \num{24} pigtails are the only solid glass in the modeled AR-HCF line.

Pigtail NLI is \emph{not} computed with the span closed form, which keeps the
self- and cross-channel mixing terms but discards the multichannel terms, in
which the two input frequencies lie in different interfering channels. The
discard is valid only when dispersion dephases those terms along the fiber. With
$\Delta\beta=4\pi^2(f_1{-}f_i)(f_2{-}f_i)[\beta_2+\pi\beta_3(f_1{+}f_2)]$, a
mixing product whose inputs lie one channel spacing from channel~$i$ accumulates
a phase mismatch $|\Delta\beta|L\approx\SI{700}{\radian}$ over an
\SI{80}{\kilo\meter} SMF span, so contributions generated at different points
along the span cancel and the discarded share stays below \SI{0.2}{\percent}.
Over a \SI{10}{\meter} pigtail the same mismatch is only \SI{0.09}{\radian} and
the loss is negligible ($\alpha L_p=\num{5e-4}$): the products add in phase
(quasi-phase matching) and are not suppressed. Pigtail NLI is therefore computed
from the full GN integral
$\tfrac{16}{27}\gamma^2B\iint G(f_1)G(f_2)G(f_1{+}f_2{-}f_i)|\rho|^2$ over every
triplet of occupied channels, with the finite-length kernel
$\rho=[1-e^{-(\alpha-j\Delta\beta)L_p}]/(\alpha-j\Delta\beta)$, stored as
coefficients of $P_jP_kP_l$ so that the load dependence is exact. The long-span
closed form would understate the band-total pigtail coefficient by
\SI{6.1}{\decibel}, least at the band edges and most mid-band; the difference
shrinks as channels drop, since a multichannel term needs three occupied
channels where a cross-channel term needs two.

\subsection{Gas Absorption}
\label{sec:gasmodel}
The AR-HCF model resolves three CO$_2$ bands near \SI{1540}{\nano\meter},
\SI{1573}{\nano\meter}, and \SI{1606}{\nano\meter} with a synthetic comb
following the gas-line model of~\cite{hcf_qamlimits2026}: \num{276}
Boltzmann-weighted Lorentzian lines spaced by the P/R-branch rotational interval
$2B\approx\SI{23.4}{\giga\hertz}$, consistent with the HITRAN line
positions~\cite{gordon2022hitran}, each of \SI{1.0}{\giga\hertz} full width at
half maximum~\cite{hcf_gasline2026}, scaled to a \SI{0.5}{\dbkm} peak
line-center loss from a measured spectrum~\cite{wang2025co2}. Only the line
spacing follows the tabulated structure; line strengths, isotopologues, pressure
shifts, Voigt broadening and the temperature and partial-pressure dependence do
not, and CO and H$_2$O are omitted, so the comb is a sensitivity spectrum rather
than a line-by-line synthesis. A residual-fill factor scales the line depth at
fixed width, so fill values are normalized notch depths, not gas concentrations.

The model treats average loss and spectral distortion separately. For channel
$i$, the loss after distance $L$ is
\begin{equation}
A_i(L) = -10\log_{10}\big\langle 10^{-g(f)L/10}\big\rangle_{f\in i},
\label{eq:gasloss}
\end{equation}
where $g(f)$ is the resolved loss profile; averaging transmitted power, rather
than loss in decibels, stops adding loss once narrow frequencies are already
dark. The equivalent coefficient $A_i(L)/L$ is included in the channel
attenuation, giving every span the same share, whereas the true increments are
front-loaded: on the worst channel the resolved spectrum gives
\SI{1.13}{\decibel} in the first span and \SI{0.26}{\decibel} in the last,
against \SI{0.40}{\decibel} assigned to each. The link total is exact, but since
the equalizer restores the spectrum after every span except the last, the
homogenized model overstates the last-span increment by \SI{0.14}{\decibel} on
the worst channel and \SI{0.06}{\decibel} in the L-band mean---more than the
residual gas penalty of Sec.~\ref{sec:gas}, which it therefore bounds rather than
predicts.

A narrow in-band notch also distorts the signal after its average power is
restored. Following~\cite{hcf_qamlimits2026}, this is represented by
\begin{equation}
P_{\mathrm{ISI}} = \Delta_{\max}\left[1-\exp\!\left(-\frac{g_i L}{A_0}\right)\right],
\label{eq:isi}
\end{equation}
where $g_i=\langle g(f)\rangle_{f\in i}$ is the excess-loss profile averaged over
channel $i$, so $g_iL$ is the accumulated in-band loss driving the notch, and
$\Delta_{\max}=\SI{5.5}{\decibel}$, $A_0=\SI{6}{\decibel}$. Since a narrow line
fills only part of the channel, $g_i$ lies about an order of magnitude below the
line-center peak, so only strongly absorbed channels approach $\Delta_{\max}$.
Both constants are inherited calibrations rather than derived quantities:
$\Delta_{\max}$ is of the order of the $Q$-factor penalties reported for deep
in-band notches at \SI{140}{\giga\baud} \num{1024}-QAM~\cite{hcf_gasline2026}.
Equation~\eqref{eq:isi} carries no symbol rate, pulse shape, modulation format,
receiver equalizer, or FEC, and, being driven by the channel-averaged loss
rather than the in-band contrast, it would also return a penalty for a
spectrally flat excess loss that cannot distort (\SI{4.4}{\decibel} for a flat
\SI{0.01}{\dbkm} over this link). It is therefore a monotone indicator of
in-band gas content with illustrative, link-specific values, not a predictive
ISI model.

\begin{table}[t]
	\centering
	\caption{System and device parameters.}
	\label{tab:params}
	\renewcommand{\arraystretch}{1.0}
	\scriptsize
	\begin{tabular}{@{}>{\raggedright\arraybackslash}p{0.60\columnwidth}
			>{\raggedright\arraybackslash}p{0.35\columnwidth}@{}}
		\toprule
		Parameter & Value \\
		\midrule
		Channels; grid; symbol rate; spans & 96; 100 GHz; 96 GBd; $12\times80$ km \\
		Extended C / L bands; guard & $2\times48$ ch, 1529.2--1566.7 / 1570.8--1610.5 nm; 400 GHz \\
		SMF $\alpha,D,S,\gamma$ & 0.21 dB/km, 17 ps/nm/km, 0.067 ps/nm$^2$/km, 1.3 /W/km \\
		AR-HCF $\alpha,D,S$ & 0.11 dB/km, 2.5 ps/nm/km, 0.005 ps/nm$^2$/km \\
		AR-HCF $n_2$ (bound); MFD; $\gamma$ (derived) & $<2.2\!\times\!10^{-23}$ m$^2$/W; 21 \si{\micro\metre}; $2.5\!\times\!10^{-4}$ /W/km \\
		Raman gain slope, silica; glass power fraction & 0.028 /W/km/THz; $10^{-4}$ \\
		IMI, AR-HCF; multipath floor SMF; TRx SNR & $-60$ dB/km; 45 dB; 26 dB \\
		Amplifier pigtail length; saturation, total out & $2\times10$ m SMF; 38 dBm \\
		Noise figure; $\eta$; $P_{\mathrm{mm}}$, C / L & 5.0 / 6.0 dB; 5.0 / 3.7 \%; 8 W \\
		Splitter loss; equalizer spacing, loss, margin & 1.5 dB; 1 span, 5 dB, 0.5 dB \\
		CO$_2$ bands; spacing; FWHM; peak & 1540/1573/1606 nm; 23.4 GHz; 1.0 GHz; 0.50 dB/km \\
		ISI $\Delta_{\max}$; $A_0$ & 5.5 dB; 6 dB \\
		Transient $\tau$; loop bandwidth; backoff & 10 ms; 500 Hz; 3 dB \\
		\bottomrule
	\end{tabular}
\end{table}
\subsection{Amplifiers, Equalization and Power}
Each span ends in a \SI{1.5}{\decibel} band splitter and separate C- and L-band
amplifiers. Under the baseline constant-gain control, amplifier $b$ compensates
the mean loss of its band:
\begin{equation}
\bar G_b=\big\langle \alpha_i L_s+\alpha^{\mathrm{gas}}_i L_s\big\rangle_{i\in b}
        + L_{\mathrm{spl}},
\label{eq:gain}
\end{equation}
where $L_s$ is the span length and $L_{\mathrm{spl}}$ is the splitter loss.
Thus only within-band variation accumulates.

A dynamic gain equalizer (DGE) restores the launch spectrum after every span
except the last in the baseline. It is a passive per-channel attenuator of
\SI{5}{\decibel} insertion loss followed by a fixed-gain stage, so restoring a
channel that sits $\delta$ below the reference needs a provisioned gain of at
least $\SI{5}{\decibel}+\delta$. The gain is sized from the largest deficit the
equalizer must correct at full occupancy (in SMF, the accumulated ISRS tilt),
plus a \SI{0.5}{\decibel} margin for partial loads: a surviving channel both
receives Raman gain from shorter-wavelength channels and loses power to
longer-wavelength ones, so dropping channels can remove more of its gain than of
its loss and leave it slightly further below the reference than at full load.
The stage's ASE and electrical power are computed at this provisioned gain,
which also caps the correction it can apply.

A spacing of $N_s$ means that no interior DGE is used.

Amplifier electrical power follows~\cite{sohanpal2026energyperbit}:
\begin{equation}
P_{\mathrm{el}}=\frac{P_{\mathrm{out}}\left(1-1/G\right)}{\eta}+P_{\mathrm{mm}},
\label{eq:power}
\end{equation}
where $P_{\mathrm{out}}$ is total optical output, $G$ the actual gain, $\eta$ the
wall-plug efficiency, and $P_{\mathrm{mm}}$ fixed monitoring power. The
efficiencies and noise figures measured for \SI{23}{\dbm}-class
devices~\cite{sohanpal2026energyperbit} are assumed to hold for the
\SI{38}{\dbm} class modeled here, constant across gain, input power and loading,
although measured efficiency varies with operating point; this extrapolation
dominates the uncertainty in the energy results. The launch ceiling is the power
at which the worst amplifier's total signal, ASE and NLI output reaches
\SI{38}{\dbm}, and a single flat per-channel launch power is chosen to maximize
ideal-Shannon throughput below it. Constant gain and constant output are the two
standard operating modes of in-line amplifiers, and constant gain is the
conventional choice in dynamic networks because constant output redistributes the
power of dropped channels to the survivors~\cite{zyskind2011book}. As the
alternative, constant-output control holds each amplifier and equalizer stage at
its full-occupancy total output, so the per-channel power rises as channels are
dropped; each equalizer still restores every channel to the reference launch and
then raises the common level to its full-load output. The resulting gains and
noise are propagated through the same cascade.

\subsection{Traffic, Load Steps and Metrics}
\label{sec:traffic}
Load sweeps use nested random channel removal, so every lower-load set is a
subset of the preceding set. Results average \numrange{20}{24} removal
orders and report the largest absolute GSNR change among the channels retained
throughout the sweep, relative to full occupancy. The energy study uses a
synthetic \SI{24}{\hour}, five-minute occupancy trace with a diurnal baseline and
five random bursts (mean occupancy \num{0.66}), a sensitivity workload;
active-channel identities are redrawn at each of the \num{288} samples, over
which both policies are averaged. The elastic policy carries only active
channels, with ASE counted only in active slots (slot-resolved filtering). The
idler policy fills vacant slots with ASE from a dedicated broadband source,
whose electrical power is modeled as \SI{5}{\watt} plus the idler optical power
divided by a \SI{2}{\percent} efficiency.

Load-step transients are calculated for the twelve line amplifiers (the interior
DGE stages are counted in the energy budget only) with the lumped single-band
reduction of the average-inversion model~\cite{sun1997avg}, which assigns a
common effective saturation energy to all channels:
\begin{equation}
\frac{\mathrm{d}g_k}{\mathrm{d}t}=\frac{g_{0,k}-g_k}{\tau}
  -\frac{P_{\mathrm{out},k}-P_{\mathrm{in},k}}{E_{\mathrm{sat}}},
\qquad P_{\mathrm{out},k}=P_{\mathrm{in},k}e^{g_k},
\label{eq:reservoir}
\end{equation}
where $g_k$ is the gain of amplifier $k$, $g_{0,k}$ its pumped gain,
$\tau=\SI{10}{\milli\second}$, and
$E_{\mathrm{sat}}=P_{\mathrm{sat}}\tau$. Because the spans are
loss-compensated, each amplifier's output change drives the next and the
response accumulates along the chain~\cite{srivastava1997transient}. Each
amplifier is assigned \SI{3}{\decibel} of saturation backoff, fixing about
\SI{2}{\decibel} of steady-state gain compression---an imposed calibration,
since intrinsic saturation power differs from rated saturated output. The pump
controller has an assumed \SI{5}{\micro\second} measurement lag and
\SI{500}{\hertz} bandwidth, a sensitivity point for which the full bandwidth
dependence is reported. A \num{200}-trial Monte Carlo varies each module's gain
($\sigma=\SI{0.5}{\decibel}$), backoff ($\sigma=\SI{1.5}{\decibel}$), lifetime
(\SI{10}{\percent} coefficient of variation), and loop bandwidth (log-normal
about \SI{500}{\hertz}). Spectral hole burning, dynamic gain tilt, ASE loading,
pump-actuator limits, and Raman-amplifier dynamics are outside this reduced
model, so the excursions are conditional model outputs rather than expected
field values.

The received GSNR is
\begin{equation}
\frac{1}{\gsnr_i}=\frac{P_{\mathrm{ASE},i}}{P_{\mathrm{sig},i}}
 +\frac{P_{\mathrm{NLI},i}}{P_{\mathrm{sig},i}}
 +\frac{1}{\mathrm{SNR}_{\mathrm{TRx}}}+\frac{1}{\mathrm{SNR}_{\mathrm{IMI}}},
\label{eq:gsnr}
\end{equation}
with all channel powers referred to the symbol-rate bandwidth. The AR-HCF IMI
floor is derived from \SI{-60}{\dbkm}~\cite{poggiolini2022hcf}, giving
\SI{30.2}{\decibel} after \SI{960}{\kilo\meter}. The penalty of
Eq.~\eqref{eq:isi} is subtracted from the GSNR in decibels. Throughput is
$\sum_i 2R_s\log_2(1+\gsnr_i)$ over active channels, and energy per bit is total
amplifier electrical power divided by throughput.
\begin{figure*}[!tb]
	\centering
	\includegraphics[width=0.57\textwidth]{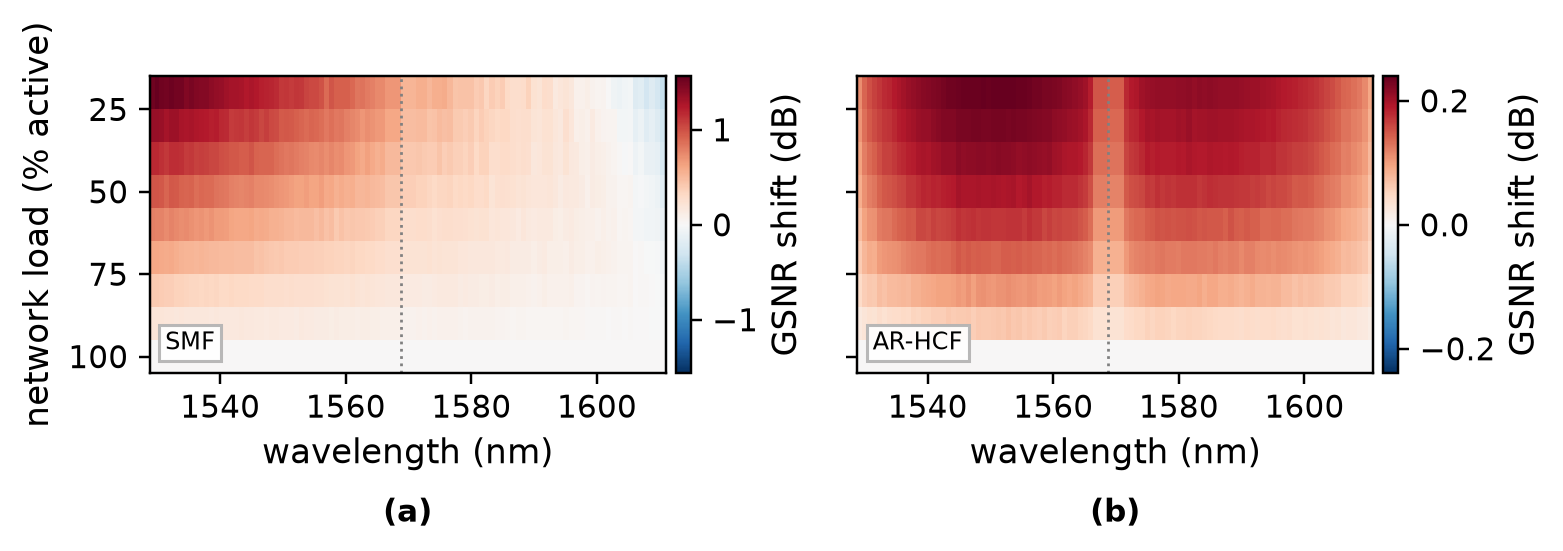}
	\caption{Per-channel GSNR change from full occupancy versus wavelength and load
		at each fiber's optimum launch power: (a) SMF and (b) AR-HCF. The color ranges
		differ; the dotted line marks the C/L guard band.}
	\label{fig:map}
\end{figure*}

\begin{figure*}[!tb]
	\centering
	\includegraphics[width=0.57\textwidth]{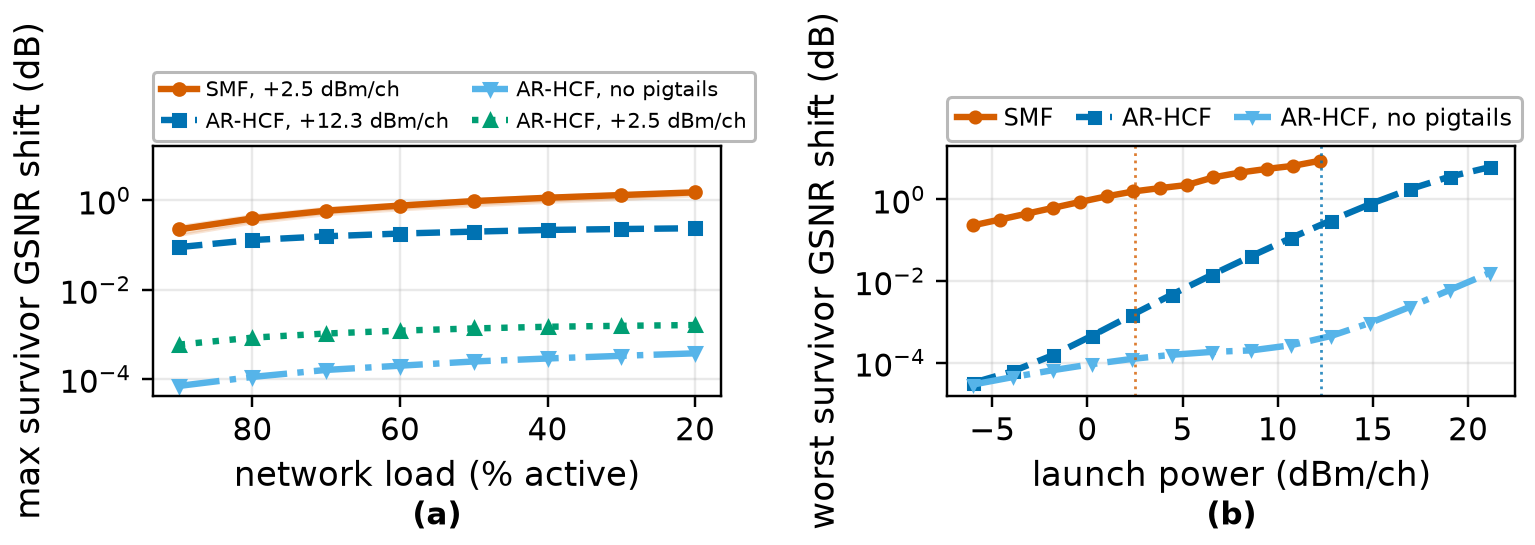}
	\caption{Largest absolute GSNR change among the channels retained throughout the
		removal sweep. (a) Versus load for SMF and for AR-HCF at the common launch, its
		own optimum, and without pigtails. Curves are means over \num{20} nested random
		removal orders. (b) Versus launch power at
		\SI{20}{\percent} occupancy, each fiber swept to its own feasibility
		ceiling; dotted lines mark the optima.}
	\label{fig:surv}
\end{figure*}
\begin{figure*}[!tb]
\centering
\includegraphics[width=0.68\textwidth]{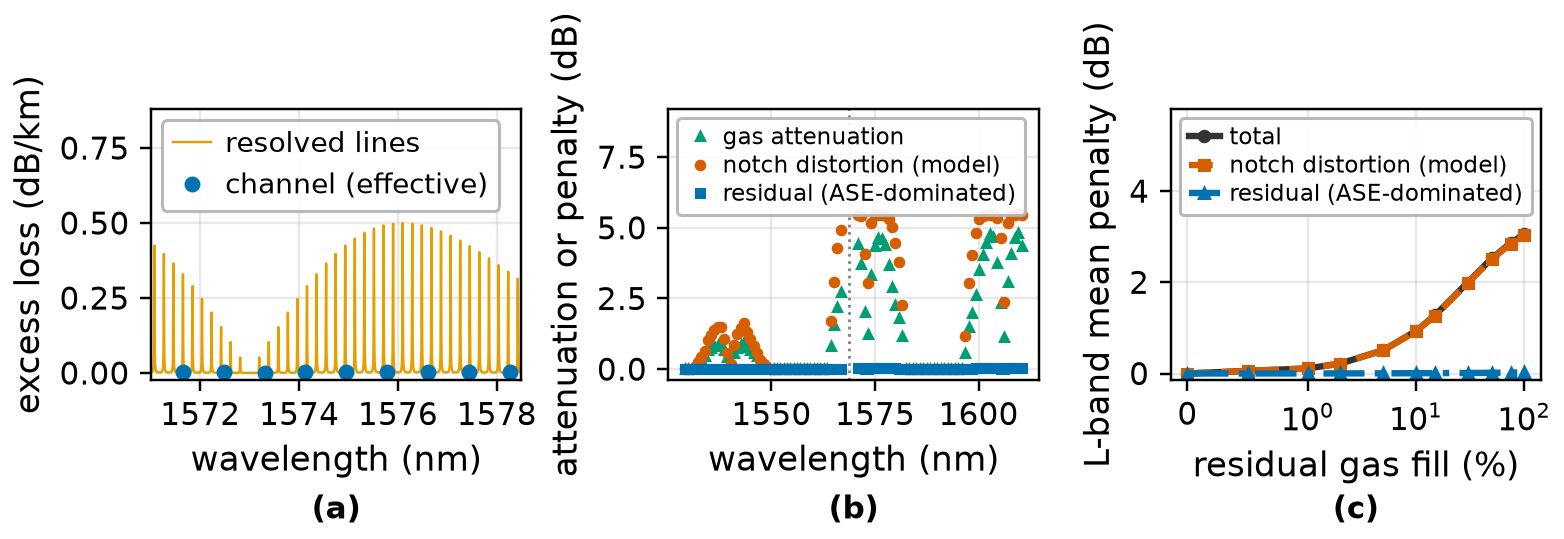}
\caption{(a) Resolved CO\textsubscript{2} lines and the effective per-channel
loss coefficient. (b) Per channel: the gas attenuation accumulated over the link,
the residual GSNR penalty left once per-span equalization has restored it
(ASE-dominated), and the notch-distortion penalty. (c) L-band mean penalty versus
residual fill. The notch-distortion curves are conditional on the assumed mapping
of Eq.~\eqref{eq:isi}, and the residual fill is a normalized notch depth rather
than a gas concentration.}
\label{fig:gas}
\end{figure*}

\section{Load Dependence of the Received GSNR}
\label{sec:load}
Fig.~\ref{fig:map} maps the change in a channel's GSNR relative to
full occupancy, averaged over \num{24} random removal orders, against wavelength
and network load, with each fiber at its own optimum launch power.
Fig.~\ref{fig:map}(a) shows that in SMF the shift grows steadily as channels are
removed and is strongly wavelength-dependent, reaching \SI{1.56}{\decibel} in the
short-wavelength region while remaining well below that in the long band. The
gradient follows the ISRS tilt: short-wavelength channels transfer power to
long-wavelength ones, so their operating point moves most when the set of Raman
pumps and interferers changes. Fig.~\ref{fig:map}(b) shows the same quantity for
AR-HCF on a color scale several times finer, peaking at
\SI{0.239}{\decibel} and nearly flat across both bands. Unlike the SMF response,
the hollow-core residual barely varies with wavelength, because it comes from
Kerr NLI generated in the pigtails, which affects all
channels alike, rather than from the wavelength-dependent ISRS transfer that tilts the
SMF shift toward the long wavelengths.

The magnitude is quantified in Fig.~\ref{fig:surv}(a), the largest GSNR shift
among the channels retained throughout the sweep against network load for four
configurations. In SMF at its optimum of $+\SI{2.5}{\dbm}$ per channel the shift
reaches \SI{1.52}{\decibel} averaged over removal orders, worst observed
\SI{1.64}{\decibel}. Operating AR-HCF at that same launch power, which isolates
the fiber from the operating point, gives \SI{1.6e-3}{\decibel}, about three
orders of magnitude lower: at this level the steady-state, fill-margin reason for
maintaining full spectral occupancy is removed, not merely reduced, though fast
transient management is not (Sec.~\ref{sec:amp}).

Operating the hollow core at its own throughput optimum of $+\SI{12.3}{\dbm}$ per
channel changes the picture. The shift rises to \SI{0.238}{\decibel}, because the
NLI that scales with the cube of the launch power is no longer
negligible. Removing the amplifier pigtails from the model lowers it to
\SI{3.8e-4}{\decibel}, so \SI{99.8}{\percent} of the residual load sensitivity at
the operating point originates in the \SI{20}{\meter} of solid glass at each
amplifier rather than in the \SI{960}{\kilo\meter} of hollow fiber. Almost all of
that pigtail contribution is generated in the output pigtail, which carries the
full amplifier output power; the input pigtail, which sees the much lower post-span
power, is negligible.

Fig.~\ref{fig:surv}(b) traces this across launch power, each fiber swept on its
own grid up to its own amplifier feasibility ceiling. The SMF curve rises
monotonically from \SI{0.23}{\decibel} at $-\SI{6}{\dbm}$ per channel to
\SI{8.89}{\decibel} at its ceiling of $+\SI{12.2}{\dbm}$. The two AR-HCF curves
coincide below about $-\SI{2}{\dbm}$ per channel, where the residual shift is set
by the fiber, and separate above it. The divergence identifies the launch power
beyond which a hollow-core line stops behaving like hollow fiber and starts
behaving like the short solid sections inside its amplifiers. Both curves run to
each fiber's feasibility ceiling, but the GN model is a first-order perturbation
result: in SMF the predicted NLI passes a tenth of the signal power near
$+\SI{7}{\dbm}$ per channel, so the upper end of that curve locates the
saturation limit rather than predicting a penalty there. Both optima lie inside
the perturbative range, at or below \SI{1}{\percent}.

Scaling the assumed couplings confirms the attribution. Removing the SMF Raman
coupling lowers its worst shift from \SI{1.60}{} to \SI{0.80}{\decibel}, so about
half of the SMF load sensitivity is ISRS-driven here. For AR-HCF,
raising the modal power fraction in glass by two orders of magnitude leaves the
worst shift essentially unchanged at \SI{0.237}{\decibel} and a further order raises
it only to \SI{0.44}{\decibel}, whereas raising the Kerr coefficient by two orders
raises it to \SI{1.64}{\decibel}: the hollow-core residual is a Kerr effect,
insensitive to the assumed Raman suppression over a wide range.

\begin{figure*}[!tb]
\centering
\includegraphics[width=0.57\textwidth]{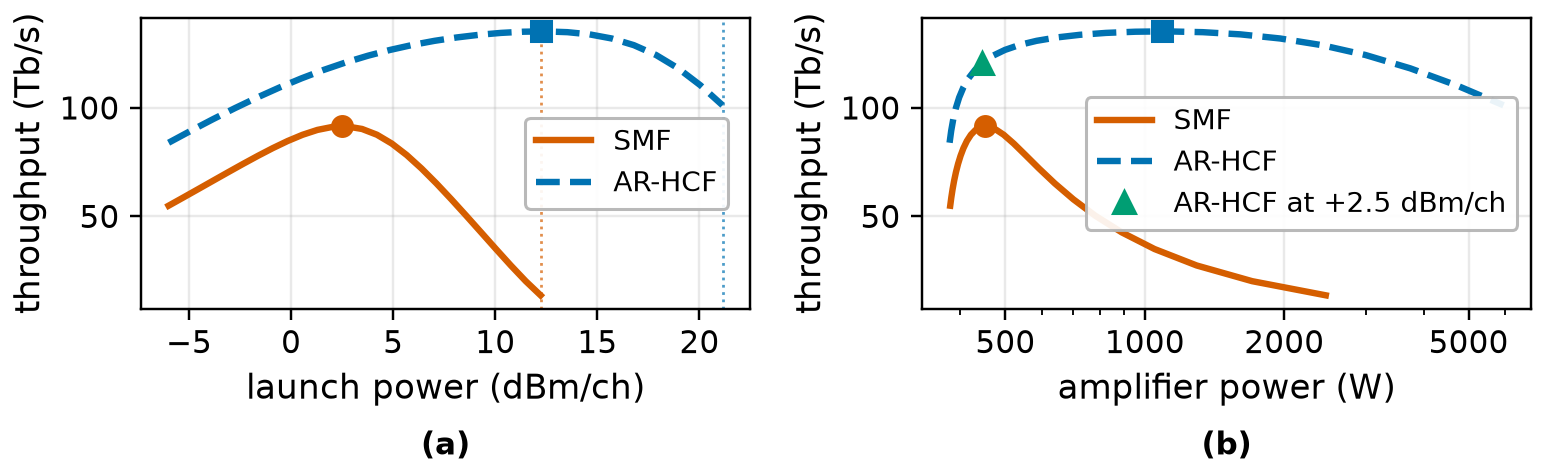}
\caption{(a) Ideal-Shannon throughput versus per-channel launch power; markers show
each optimum and dotted lines the amplifier feasibility ceilings. (b) The same
sweeps as throughput versus amplifier power, with the AR-HCF point at the SMF
launch power marked separately.}
\label{fig:launch}
\end{figure*}

\section{Gas-Line Absorption}
\label{sec:gas}
The resolved absorption profile around the strongest carbon-dioxide band is shown
in Fig.~\ref{fig:gas}(a), together with the effective loss each channel
experiences over the link. The line center reaches \SI{0.5}{\dbkm}, and the
arithmetic mean over a \SI{96}{\giga\hertz} channel is at most \SI{0.037}{\dbkm}
in the L band and \SI{0.014}{\dbkm} in the C band, of the same order as the
\SIrange{0.03}{0.08}{\dbkm} reported from deployed
fiber~\cite{hong2026tpo}.
Propagated through Eq.~\eqref{eq:gasloss}, the
physical gas attenuation of a channel over the \SI{960}{\kilo\meter} link is
substantial: it averages \SI{2.1}{\decibel} across the L band and reaches
\SI{4.8}{\decibel} on the worst channel.

Once that loss is compensated, the two contributions to the GSNR penalty are
shown channel by channel in Fig.~\ref{fig:gas}(b). The residual propagation
penalty is small---about \SI{0.024}{\decibel} average across the L band and at
most \SI{0.04}{\decibel}---because the per-span equalizer restores the
compensable level for only a little extra ASE, \SI{0.0225}{\decibel} of the
L-band mean, the remainder being the small change in NLI the altered power
profile brings with it. It is not the physical gas loss above, and as
Sec.~\ref{sec:gasmodel} notes it is smaller than the longitudinal uncertainty of
the homogenized gas model, so it bounds the residual rather than resolving it.
The notch distortion, under the assumed bounded mapping of Eq.~\eqref{eq:isi}, varies
strongly from channel to channel with the in-band gas content. In the C band it
averages about \SI{0.7}{\decibel}, reaching \SI{1.6}{\decibel} on the channels
sitting on the \SI{1540}{\nano\meter} lines and \SI{4.9}{\decibel} on the single
channel at the C/L edge, \SI{1566.7}{\nano\meter}, which catches the wing of the
\SI{1573}{\nano\meter} band, so the small C-band average is not a bound on
the worst C-band channel. In the L band, where
the strong 1573 and 1606 nm combs fall in-band, it averages about
\SI{3}{\decibel} and approaches the \SI{5.5}{\decibel} asymptote on the worst
channels-on-line. Because Eq.~\eqref{eq:isi} carries no symbol rate, pulse shape, or
receiver equalizer, its magnitude is illustrative only: it indicates that a narrow
in-band null costs more than the small average power it removes.

This band asymmetry is externally corroborated. A waveform-level study of C$+$L
transmission over \SI{1000}{\kilo\meter} likewise reports limited
C-band but severe L-band degradation, with ideal gas suppression adding
\SI{17.9}{\tbps} of average L-band throughput~\cite{gan2026gla}. Its sensitivity
to gas content is the stronger one: reducing CO$_2$ from \num{1000} to
\SI{400}{ppm} recovers \SI{2.3}{\decibel} of SNR on average and
\SI{4.9}{\decibel} at worst on the affected L-band channels after one
\SI{100}{\kilo\meter} span, against \SI{0.8}{} and \SI{1.3}{\decibel} for the same
fractional reduction applied to Eq.~\eqref{eq:isi}. The systems differ, so the
values are not directly comparable, but Eq.~\eqref{eq:isi} understates how
sharply the impairment responds to gas removal.

Fig.~\ref{fig:gas}(c) illustrates that, under this mapping, the modelled distortion
falls with residual fill---about \SI{0.9}{\decibel} remains at \SI{10}{\percent} fill
and it drops below \SI{0.2}{\decibel} only near \SI{1}{\percent}. The residual
curve, by contrast, stays near the axis throughout: it is the same post-equalization
residual as in Fig.~\ref{fig:gas}(b), and although it too rises with fill it
reaches only a few hundredths of a decibel at full fill, because a flat-gain
amplifier compensates the gas loss but cannot fill the narrow notch. The
distortion therefore falls only slowly as gas is removed, so gas
reduction is useful but no required residual level follows from these curves:
the fill at which it becomes negligible is a property of the assumed
mapping parameters and of the synthetic line depth, and the values are read as
trends, not specifications.

Gas removal by evacuation, purging or preform
pretreatment~\cite{xiong2025co2elim} is one route to reduce the notch impact, but
purging becomes slow beyond
a few kilometres and is not yet practical over long links. The impairment can
also be mitigated in the digital domain: spectral
pre-equalization recovers
several decibels of $Q$ over a short receiver equalizer for \SI{140}{\giga\baud}
\num{1024}-QAM~\cite{hcf_gasline2026} and spectrum-optimized sub-carrier
multiplexing improves BER by more than an order of
magnitude~\cite{wang2025co2}, with field
measurements confirming the system dependence~\cite{ge2025fieldtrial}.

\section{Launch Power and Amplifier Energy}
\label{sec:amp}

\subsection{Launch-Power Optimum}
Fig.~\ref{fig:launch}(a) shows the total ideal-Shannon throughput against
per-channel launch power, each curve terminated at its own amplifier feasibility
ceiling. SMF peaks at $+\SI{2.5}{\dbm}$ per channel and \SI{91.5}{\tbps}, well
below its ceiling of $+\SI{12.2}{\dbm}$, so its operating point is set by
nonlinearity and not by saturation. AR-HCF peaks at $+\SI{12.3}{\dbm}$ and
\SI{135.3}{\tbps}, also below its ceiling of $+\SI{21.1}{\dbm}$: a launch power
\SI{9.8}{\decibel} higher and \SI{1.5}{\times} the throughput, but a nonlinear
optimum nonetheless.

That optimum is set almost entirely by the pigtails. At $+\SI{12.3}{\dbm}$ per
channel, \SI{99.8}{\percent} of the NLI reaching the receiver is generated in the
\SI{20}{\meter} of solid pigtail at each amplifier, almost all in the output pigtail
carrying the full output, rather than in the \SI{960}{\kilo\meter} of hollow fiber.
Removing them leaves the line limited by amplifier saturation rather than by
nonlinearity at all---its throughput still rises at its ceiling of
$+\SI{21.2}{\dbm}$ per channel---so the pigtails hold the optimum
\SI{8.9}{\decibel} below the launch ceiling. That launch margin carries little
throughput, however. Without pigtails the line delivers \SI{136.6}{\tbps} at the
same $+\SI{12.3}{\dbm}$ per channel, only \SI{1.3}{\tbps} more, and
\SI{138.8}{\tbps} at its ceiling for \num{5.5} times the amplifier power, because
the GSNR is already close to the limit set by the transceiver and IMI floors and
the gas-notch penalty. The pigtails therefore dominate the NLI and the load
sensitivity, but not the throughput. Their NLI scales with the number
of amplifier sites rather than distance, a component-count constraint consistent
with the nonlinear threshold of a hybrid hollow-core/solid span being set by the
solid segment~\cite{mardoyan2026hybridspan}.

Fig.~\ref{fig:launch}(b) recasts the sweeps as throughput against amplifier power,
launch power being the parameter along each curve. At the common launch power of
$+\SI{2.5}{\dbm}$ per channel, marked separately, AR-HCF delivers \SI{120.5}{\tbps}
for \SI{444}{\watt} against \SI{91.5}{\tbps} for \SI{453}{\watt} in SMF, so
\SI{1.3}{\times} the throughput for the same power, at \SI{3.69}{} against
\SI{4.94}{\pico\joule\per\bit}. At its own optimum AR-HCF delivers \SI{135.3}{\tbps}
but draws \SI{1089}{\watt}, \SI{8.05}{\pico\joule\per\bit} or \SI{1.6}{\times} the
SMF figure. The additional \SI{14.8}{\tbps} from moving to the hollow-core optimum
costs a factor of \num{2.4} in amplifier power, so whether it is worth taking
depends on whether throughput or energy binds.

\subsection{Amplifier Energy and Load Transients}
\begin{figure*}[!tb]
\centering
\includegraphics[width=0.74\textwidth]{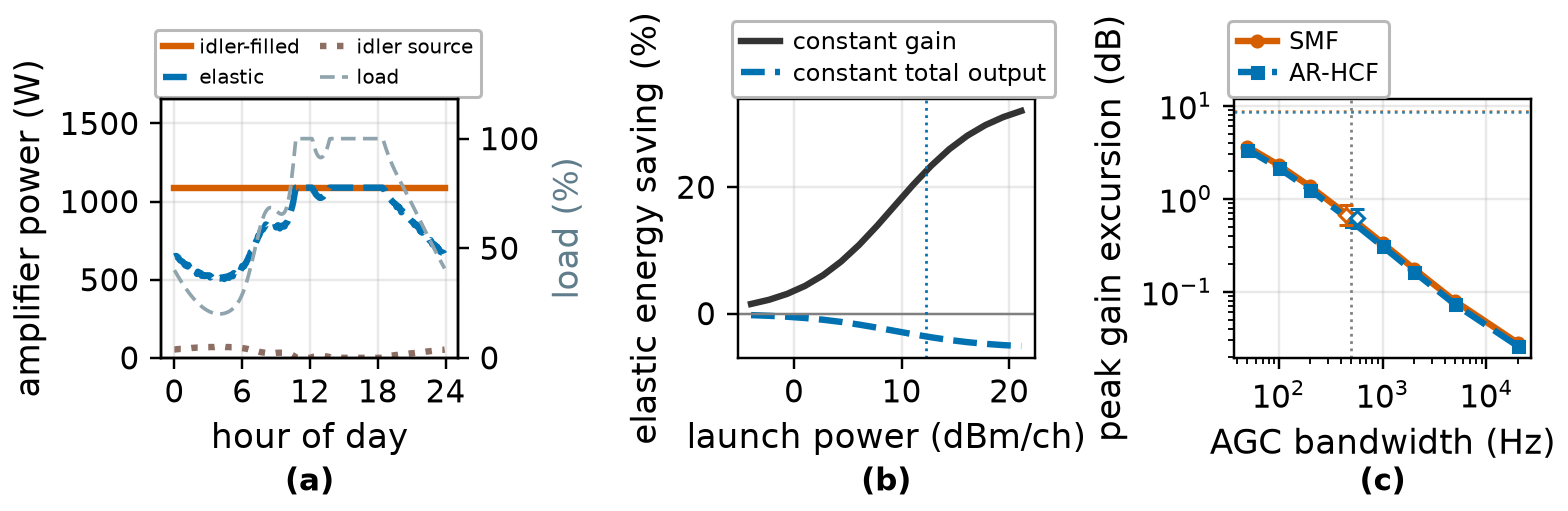}
\caption{(a) Amplifier power of the AR-HCF line at its optimum launch power under
the idler-filled and elastic policies, with the \SI{24}{\hour} occupancy trace as
the dashed grey curve (right axis). (b) Day-averaged elastic energy saving versus
per-channel launch power for two gain-control policies; the dotted line marks the
optimum. (c) Peak gain excursion after a load step versus gain-control bandwidth.
Dotted lines are the uncontrolled asymptotes; open symbols give the median and
\numrange{5}{95}th percentiles over \num{200} chains of non-identical modules.}
\label{fig:energy}
\end{figure*}
 Fig.~\ref{fig:energy}(a) shows the electrical power of the AR-HCF
amplifiers over the \SI{24}{\hour} occupancy trace at the
throughput-optimal launch power. Under the idler policy, vacant slots are filled
with ASE, holding the amplifier loading and electrical draw at
\SI{1089}{\watt}; under the elastic policy only active channels are amplified,
so the power follows the traffic and the day-averaged
consumption falls by \SI{22.6}{\percent}. That percentage
compares amplifier power only; the broadband ASE source that generates the idler
light (Sec.~\ref{sec:traffic}) accounts for a further \SI{2.8}{\percent} of the
total idler-policy draw, amplifiers plus source.

The modeled link contains \num{24}
line-amplifier stages (\num{12} spans times two bands) and \num{22} amplified-DGE
stages, \num{46} in total; at \SI{8}{\watt} each~\cite{sohanpal2026energyperbit}
their fixed contribution is \SI{368}{\watt}, leaving
\SI{721}{\watt} of the full-load total in the load-dependent amplification
term. Under constant gain that term behaves as expected for a gain-controlled
amplifier: its output, and with it the pump term of Eq.~\eqref{eq:power}, scales
with the number of channels carried~\cite{zyskind2011book}, so the amplifier
power that can be shed scales with the fraction of the band left \emph{vacant},
and the day-averaged saving is close to the product of the mean
vacant fraction, $1-0.66=0.34$, and the pump-dependent share of
\SI{66}{\percent}, that is \SI{22.5}{\percent}. It is therefore set
by the traffic trace and by the fixed-to-pump split of Eq.~\eqref{eq:power}, not
by the fiber; since monitoring and management electronics can dominate
amplifier consumption~\cite{lundberg2017power}, a fixed overhead above the
\SI{8}{\watt} per stage assumed here would lower the saving with the pump share.
The absolute wattage is likewise a property of the
assumed line architecture, in which a separately powered amplified equalizer
follows every span: with no interior equalizer the same link at the same launch
power draws \SI{616}{\watt} instead of \SI{1089}{\watt} and delivers
\SI{135.6}{\tbps} instead of \SI{135.3}{\tbps}, the modeled equalizer being ideal
in flatness, while the constant-gain elastic saving rises by less than a point,
to \SI{23.5}{\percent}. The
percentages and their control-policy dependence therefore transfer more readily
than the absolute powers, which exclude terminal equipment,
site cooling, and network-control hardware; energy per bit uses
the ideal Shannon throughput.

Fig.~\ref{fig:energy}(b) shows how the day-averaged saving depends on launch
power and on the control policy. Under constant-gain control it grows from
\SI{1.5}{\percent} at
$-\SI{4}{\dbm}$ per channel to \SI{32.1}{\percent} at the feasibility ceiling,
because the load-dependent pump term of Eq.~\eqref{eq:power}
grows relative to the fixed monitoring power: at low launch power almost all the
draw is the \SI{368}{\watt} of fixed overhead, which no reduction in channel count
can lower, whereas near the optimum the pump term dominates and carrying fewer
channels lowers it in proportion. Under constant-output control the saving
vanishes, as it must: the pump term of Eq.~\eqref{eq:power} is fixed by the
output, which this policy holds at its full-load value, leaving only the factor
$1-1/G$ free to change. Solved on the same cascade, the day-averaged saving at
the optimum is $-\SI{3.6}{\percent}$. Only \SI{1.2}{\watt} of the
\SI{39}{\watt} increase arises in the line amplifiers, all of it at the first
amplifier of each band: holding the total output holds the total input of every
later amplifier at its full-load value, so their gains barely change. The other
\SI{97}{\percent} arises in the equalizer stages, because the modeled equalizer
restores each channel to the reference launch before raising the level back to
its full-load output, discarding and re-amplifying at every span the
per-channel power gained upstream. Without interior equalizers the
constant-output saving is $-\SI{0.2}{\percent}$, so the negative value belongs
to the equalizer model and the robust result is the absence of a saving.
Neither outcome is a new amplifier effect; together they quantify, for this link, how much of the energy freed by
elastic operation the control policy lets the line keep, confirming that the
benefit belongs to the control strategy rather than to the fiber.

Fig.~\ref{fig:energy}(c) shows the peak gain excursion accumulated over the chain
of twelve line amplifiers when \SI{85}{\percent} of the channels are dropped at
once, a stress step chosen for sensitivity rather than taken from the
traffic trace. Uncontrolled, the
excursion reaches \SI{8.77}{\decibel} in SMF and \SI{8.58}{\decibel} in AR-HCF,
consistent with the several-decibel excursions reported for uncontrolled
cascades~\cite{srivastava1997transient}. With gain control the residual is set by
the loop bandwidth, falling roughly as its inverse: for the AR-HCF L band,
\SI{3.37}{\decibel} at
\SI{50}{\hertz}, \SI{0.57}{\decibel} at \SI{500}{\hertz} and \SI{0.026}{\decibel}
at \SI{20}{\kilo\hertz}, with the SMF L band within \SI{0.25}{\decibel}
throughout. The model reaches a sub-decibel residual for loop
bandwidths of a few hundred hertz. The open symbols in Fig.~\ref{fig:energy}(c)
show the sensitivity to differences
among amplifier modules at the nominal \SI{500}{\hertz} bandwidth, over the
\num{200} realizations of Sec.~\ref{sec:model}. The median peak excursion
is \SI{0.67}{\decibel} for SMF and \SI{0.62}{\decibel} for AR-HCF, with
5th--95th-percentile intervals of \SIrange{0.53}{0.85}{\decibel} and
\SIrange{0.48}{0.78}{\decibel}, respectively. The identical-module predictions,
\SI{0.62}{\decibel} and \SI{0.57}{\decibel}, lie slightly below these medians.

The small separation between the two curves reflects the scope of the dynamic
model: both links use the same \SI{3}{\decibel} saturation backoff and so have
nearly the same steady-state gain compression, leaving their average-gain
reservoir responses about \SI{0.05}{\decibel} apart at \SI{500}{\hertz}. A
deployed wideband link can also exhibit spectral-hole burning, dynamic gain tilt,
fiber SRS and more detailed controller dynamics; AR-HCF suppresses the
transmission-fiber SRS contribution, but not the amplifier-related ones.

This also clarifies the role of ASE idlers. An instantaneous, equal-power ASE
replacement leaves the total amplifier input unchanged and so produces no
excursion in the scalar reservoir model, but real replacement has finite
detection and switching delays and may differ from the dropped traffic in power
or spectral shape, leaving a residual transient. Experimentally, automated filler
ASE reduced the required line-amplifier adjustments from \num{26} to \num{3}
during channel loading, while protection ASE shortened failure recovery by more
than \SI{90}{\percent} in a \SI{1875}{\kilo\meter} C$+$L
link~\cite{aseidler2024}. Idlers therefore mitigate disturbances rather than
guaranteeing a transient-free event: AR-HCF removes the modeled steady-state
Kerr/ISRS rationale for continuous filling, but transient events must still be
managed by coordinated ASE replacement, amplifier gain control, or both.

\section{Conclusion}
  The performance and energy efficiency of AR-HCF links
under variable traffic depend on the complete transmission system, not on the
hollow-core fiber alone. With Kerr and ISRS coupling in the transmission fiber
greatly reduced, the remaining load dependence is governed almost entirely by the
high-power solid-glass sections at the amplifier sites---whose nonlinearity has
to be evaluated outside the long-span approximation---so shortening or replacing
them, and coordinating amplifier control with traffic, are the more immediate
priorities. Continuous ASE filling is not required by the modeled steady-state
propagation behavior, but ASE replacement and gain control remain complementary
options for managing add/drop transients. As in any amplified line, traffic-proportional energy consumption is achievable only when amplifier control allows electrical power to scale with the carried optical load. Constant-gain operation enables this behavior, whereas constant-output-power operation does not. Thus, it is not an intrinsic consequence of changing the fiber type.

\bibliographystyle{IEEEtran}
\bibliography{refs_v10}

\end{document}